\documentclass{article}

\usepackage{arxiv}

\usepackage[utf8]{inputenc}
\usepackage[T1]{fontenc}
\usepackage{hyperref}
\usepackage{url}
\usepackage{amsfonts}
\usepackage{nicefrac}
\usepackage{microtype}
\usepackage{lipsum}
\usepackage{graphicx} 
\usepackage{xcolor}
\usepackage{amsmath}
\usepackage{amssymb}
\usepackage{subcaption}

\title{Optimization-based Unfolding in High-Energy Physics}

\author{
    Simone Gasperini$^{1,2}$,
    Gianluca Bianco$^{1,2}$,
    Marco Lorusso$^{3}$,
    Carla Rieger$^{4,5}$,
    Michele Grossi$^{5}$\\[0.5em]
    $^{1}$Dipartimento di Fisica e Astronomia (DIFA), University of Bologna, Italy\\
    $^{2}$Istituto Nazionale di Fisica Nucleare (INFN), Sezione di Bologna, Italy\\
    $^{3}$Istituto Nazionale di Fisica Nucleare (INFN), CNAF, Italy\\
    $^{4}$School of Engineering and Design, Technical University of Munich (TUM), Germany\\
    $^{5}$European Organization for Nuclear Research (CERN), Geneva, Switzerland
}

\begin{document}
\maketitle

\begin{abstract}
    In experimental High-Energy Physics, unfolding refers to the problem of estimating the underlying distribution of a physical observable from detector-level data, in the presence of statistical fluctuations and systematic uncertainties.
    Starting from its reformulation as a regularized quadratic optimization problem, we develop a framework to address unfolding using both classical and quantum-compatible methods.
    In particular, we derive a Quadratic Unconstrained Binary Optimization (QUBO) representation of the unfolding objective, allowing direct implementation on quantum annealing and hybrid quantum-classical solvers.
    The proposed approach is implemented in QUnfold, an open-source Python package integrating classical mixed-integer solvers and D-Wave's hybrid quantum solver.
    We benchmark the method against widely used unfolding techniques in RooUnfold, including response Matrix Inversion, Iterative Bayesian Unfolding, and Singular Value Decomposition unfolding, using synthetic datasets with controlled distortion effects.
    Our results demonstrate that the optimization-based approach achieves competitive
    reconstruction accuracy across multiple distributions while naturally accommodating regularization within the objective function.
    This work establishes a unified optimization perspective on unfolding and provides a practical pathway for exploring quantum-enhanced methods in experimental HEP data analysis.
\end{abstract}

\section{Introduction}\label{sec:intro}
In High-Energy Physics, vast amounts of experimental data are collected and analyzed to extract meaningful physical insights, validate theoretical models, and search for potential new phenomena~\cite{fruhwirth2000data, blobel2002unfolding, brun1997root}.
However, the observable distributions measured by particle detectors do not directly correspond to the underlying physical truth.
Finite detector resolution, limited acceptance, inefficiencies, electronic noise, and reconstruction effects distort the observed quantities~\cite{knoll2010radiation}.
In addition, statistical fluctuations and background processes with similar signatures further modify the measured spectra~\cite{schmitt2017data}.
As a result, the comparison between experimental data and theoretical predictions requires correcting for detector effects in a statistically consistent manner.

Statistical unfolding refers to the set of techniques used to obtain an estimate of the underlying true distribution of a physical observable from detector-level measurements, accounting for detector effects and associated uncertainties.
Mathematically, unfolding constitutes an ill-posed inverse problem that is often formulated as a deconvolution task~\cite{cormier_unfolding_2019}.
Small statistical fluctuations in the measured data may lead to large variations in the reconstructed solution, making naive inversion techniques unstable.
Regularization strategies are therefore essential to stabilize the solution while controlling bias.
Unfolding plays a central role in precision measurements and searches at collider experiments, as it enables model-independent comparisons between theory and experiment~\cite{han2024quantum}.

Several unfolding techniques are routinely employed in HEP analyses and are implemented in frameworks such as \texttt{RooUnfold}\footnote{\url{https://gitlab.cern.ch/RooUnfold/RooUnfold}}. 
Widely used approaches include Singular Value Decomposition (SVD), Iterative Bayesian Unfolding (IBU), and related regularized matrix-based methods.
Although matrix inversion is formally possible, it is rarely applied in practice due to its strong sensitivity to statistical noise and the typically ill-conditioned nature of detector response matrices~\cite{2020_comparison}.
More recently, machine-learning-based unfolding strategies have been proposed to improve flexibility and reduce bias~\cite{andreassen2020omnifold, huetsch2024landscape, backes2024event}.
Despite these advances, unfolding remains a challenging task, especially when balancing statistical precision, regularization strength, and computational efficiency.

In this work, we adopt a complementary perspective and reformulate the unfolding problem as a discrete optimization task.
Starting from a regularized likelihood formulation, we derive an equivalent quadratic integer optimization model that naturally incorporates detector response and smoothness constraints.
This viewpoint enables the use of modern optimization techniques, including mixed-integer quadratic programming and quantum-compatible formulations.
In particular, we construct a Quadratic Unconstrained Binary Optimization (QUBO) representation of the unfolding objective, allowing the problem to be mapped onto quantum annealing hardware.

To facilitate practical applications, we developed \texttt{QUnfold}\footnote{\url{https://github.com/Quantum4HEP/QUnfold}}, an open-source Python package that integrates classical solvers, hybrid quantum-classical solvers, and quantum annealing backends within a unified framework.
Our approach builds upon previous proof-of-principle studies employing quantum annealing for unfolding~\cite{cormier_unfolding_2019}, extending them to a more general and scalable setting suitable for systematic benchmarking.
Unlike earlier exploratory studies, we provide a comprehensive comparison against established methods, evaluating reconstruction accuracy and stability across multiple distribution types.

To assess performance, we evaluate the proposed framework on several artificially generated distributions with controlled detector effects.
The results demonstrate that the optimization-based formulation achieves competitive reconstruction quality compared to standard methods.
Moreover, the QUBO representation provides a direct pathway for exploring quantum-enhanced optimization techniques within realistic HEP analysis workflows~\cite{di2024quantum, franchini_2025}.

\section{Preliminaries}\label{sec:preliminary}
In HEP experiments, the folding process describes how the true distribution of a physical observable is transformed by detector effects into the measured spectrum.
As particles traverse the detector, their reconstructed signals are affected by finite resolution, limited acceptance, inefficiencies, and electronic noise.
Consequently, the measured distribution differs from the underlying physical truth.
All experimentally measured observables are subject to this transformation, which is intrinsic to the detection process.

\subsection{Unfolding problem definition}
The goal of statistical unfolding is to estimate the true probability density function $f(z)$ of a physical observable $z$ from the corresponding detector-level distribution $g(\mu)$ of the reconstructed observable $\mu$~\cite{cowan1998statistical}.
The detector effects are modeled by the response function $r(\mu|z)$, which represents the conditional probability of observing $\mu$ given a true value $z$.
This function can be factorized into a migration (resolution) component $m(\mu|z)$ and a detection efficiency $\epsilon(z)$:
\begin{equation}
    r(\mu|z) = m(\mu|z)\,\epsilon(z) \, .
    \label{eqn:prob_r}
\end{equation}

In practice, the response function is usually estimated using Monte Carlo simulations of the detector and reconstruction chain~\cite{blobel2002unfolding}.
The expected detector-level distribution is then given by
\begin{equation}
    g(\mu) = \int r(\mu|z) f(z)\, dz + \beta(\mu) \, ,
    \label{eqn:func_g}
\end{equation}
where $\beta(\mu)$ denotes the expected background contribution.
Recovering $f(z)$ from $g(\mu)$ constitutes an ill-posed inverse problem and it corresponds to a deconvolution task.

In experimental analyses, distributions are discretized into histograms with $M$ bins and the continuous density functions are replaced by integer-valued vectors.
We denote the true histogram and measured histogram by $\boldsymbol{z}$ and $\boldsymbol{n}$ respectively, where $z_j$ and $n_j$ are the number of actual and detected events in bin $j$
The response function is represented by the response matrix $R \in \mathbb{R}^{M \times M}$, where $R_{ij}$ is the probability that an event generated in true bin $j$ is reconstructed in measured bin $i$.
The discrete version of Eq.~\eqref{eqn:func_g} becomes
\begin{equation}
    \mu_i = \sum_{j=1}^{M} R_{ij} z_j + \beta_i \, .
    \label{eqn:folding_R}
\end{equation}

\subsection{State-of-the-art methods}
The most direct way to solve Eq.~\eqref{eqn:folding_R} is through matrix inversion~\cite{cowan1998statistical}.
In practice, however, one has access to the measured histogram $\boldsymbol{n}$, rather than to the expected reconstructed $\boldsymbol{\mu}$.
Replacing $\boldsymbol{\mu}$ with the observed $\boldsymbol{n}$ and inverting the relation leads to the Matrix Inversion (MI) estimator
\begin{equation}
    \hat{\boldsymbol{z}} = R^{-1}(\boldsymbol{n} - \boldsymbol{\beta}) \, .
    \label{eqn:matrix-inv}
\end{equation}
This method is highly sensitive to statistical fluctuations because response matrices are typically ill-conditioned~\cite{hoecker1996svd}.
Small perturbations in the data may therefore induce large oscillations in the unfolded result.
For this reason, direct inversion is rarely used in realistic analyses~\cite{d1995multidimensional}.

Modern unfolding techniques stabilize the inverse problem through regularization.
Some approaches introduce explicit penalty terms, such as Tikhonov regularization~\cite{tikhonov1963solution}, while others rely on iterative procedures with implicit regularization via early stopping.
A widely used method is based on the Singular Value Decomposition (SVD) of the response matrix~\cite{hoecker1996svd}.
The SVD decomposes $R$ into orthogonal components, allowing small singular values to be suppressed reducing noise.
The regularization strength is controlled by truncating the number of retained components.
Another standard approach is Iterative Bayesian Unfolding (IBU)~\cite{d1995multidimensional}.
Starting from an initial prior distribution $\hat{\boldsymbol{z}}^{(0)}$, the method iteratively updates the estimate using Bayes theorem, forming an Expectation-Maximization (EM) procedure~\cite{dempster1977maximum}.
In practical applications, the iteration is stopped after a finite number of steps rather than at full convergence.
This early stopping acts as an implicit regularization mechanism that mitigates overfitting to statistical noise.
Both SVD and IBU are commonly applied in HEP data analyses through the \texttt{RooUnfold} framework.

\subsection{Log-likelihood maximization}
An alternative and statistically well-motivated framework formulates unfolding as a likelihood maximization problem~\cite{cowan1998statistical}.
Instead of inverting the response matrix, one estimates the underlying true distribution $\boldsymbol{z}$ that maximizes the probability of observing the measured data $\boldsymbol{n}$ given the detector response.

Assuming Poisson-distributed counts in each bin, the log-likelihood can be written as
\begin{equation}
    \log \mathcal{L}(\boldsymbol{z}) 
    = \sum_{i=1}^{M} \log \mathcal{P}(n_i; \mu_i) \, ,
    \label{eqn:poisson}
\end{equation}
where $\mu_i$ is given by Eq.~\eqref{eqn:folding_R}.
To stabilize the solution, a regularization term is added:
\begin{equation}
    \max_{\boldsymbol{z}} 
    \left( \log \mathcal{L}(\boldsymbol{z}) + \lambda S(\boldsymbol{z}) \right) \, .
    \label{log-like-max}
\end{equation}
The parameter $\lambda$ controls the strength of the regularization.
Choosing $\lambda$ appropriately is crucial to balance variance and bias.
A common smoothness prior penalizes large curvature of the unfolded distribution.
In the continuous limit, this can be expressed as
\begin{equation}
    S[f(z)] = - \int \left( \frac{d^k f(y)}{dy^k} \right)^2 dy \, ,
    \label{eqn:reg}
\end{equation}
where typically $k=2$.
In the discrete case, this corresponds to a Laplacian penalty
\begin{equation}
    S(\boldsymbol{z}) 
    = - \sum_{i=1}^{M-2} 
    (-z_i + 2z_{i+1} - z_{i+2})^2 \, .
    \label{eqn:laplace}
\end{equation}

Likelihood-based unfolding provides a statistically consistent framework and allows explicit control over regularization. Within this framework, the IBU method can be viewed as an EM algorithm for a Poisson likelihood and, in the limit of infinitely many iterations, converges to the unregularized maximum-likelihood solution~\cite{Wu1983}.

\section{Methods}\label{sec:methods}
In this section, we present an optimization-based reformulation of the statistical unfolding problem in~\eqref{log-like-max}.
We outline several strategies for addressing the task in the simpler scenario of no background (corresponding to the case of $\beta = 0$ in~\eqref{eqn:matrix-inv}), including the possibility of framing it within a quantum computing framework.

We first introduce a quadratic optimization model that is well suited to classical heuristic approaches for computing a sub-optimal solution.
We then reformulate this model as a Quadratic Unconstrained Binary Optimization (QUBO) problem, enabling its direct solution via Quantum Annealing on dedicated quantum hardware.

\subsection{Quadratic optimization formulation}\label{sec:quadratic_opt}
Starting from the regularized log-likelihood maximization in~\eqref{log-like-max}, we reformulate the unfolding problem as a discrete optimization-based task following the same approach presented in~\cite{cormier_unfolding_2019}.
First, we replace the sum over the Poisson terms in~\eqref{eqn:poisson} by the squared L2-norm of the difference between the expected and the measured histograms.
This corresponds to taking the Gaussian approximation of the Poisson distribution for each bin in the limit of a large number of entries.
Then, using finite differences to approximate the 2\textsuperscript{nd}-order derivative as described in~\eqref{eqn:laplace}, we write the regularization term as the squared L2-norm of the discrete Laplacian operator $D$ applied to the unfolded histogram~$\boldsymbol{z}$, up to the multiplicative factor $\lambda$.
Thus, the equivalent optimization-based version of the unfolding problem reads:
    \begin{equation}
        \label{l2-norm-min}
        \min_{\boldsymbol{z}} \left( ||R \boldsymbol{z} - {\boldsymbol{n}}||^2 + \lambda ||D \boldsymbol{z}||^2 \right) \, .
    \end{equation}

Quadratic optimization problems like this one are challenging as they might be NP-hard, considering, for example, the Quadratic Knapsack problem introduced in~\cite{gallo1980quadratic}.
Solution strategies range from exact methods such as branch-and-bound and semi-definite programming to classical heuristics~\cite{burer2003nonlinear, korner1985integer}.
Given the formulation as a quadratic optimization problem in~\eqref{l2-norm-min},
classical solvers can be employed to find a sub-optimal solution representing the unfolded histogram $\Hat{\boldsymbol{z}}$.

In this work, we use the commercial software Gurobi to address this optimization problem.
This solver offers the state-of-the-art mathematical optimization tools for linear and mixed-integer programming.
In addition to the classical optimization approach, we employ D-Wave's hybrid quantum-classical solver, which generates samples of candidate solutions to optimization problems defined over integer variables.
The central idea of a hybrid quantum-classical model is to combine the complementary strengths of classical and quantum computing resources.
Classical processors are used for tasks such as problem decomposition, constraint management, and coordination of the optimization workflow~\cite{osaba2025dwave}.
Quantum processors, in contrast, are designed to explore complex and high-dimensional energy landscapes and may leverage quantum effects to traverse barriers between local minima.
This collaborative approach aims to improve the exploration of the solution space and support the search for near-optimal solutions in challenging nonconvex settings.
Through the hybrid structure of this D-Wave solver, complex problems beyond the capacity of standalone quantum processors can be tackled~\cite{DWave2020HybridSolver}.

\subsection{QUBO model formulation}\label{sec:qubo_formulation}
To employ quantum-based optimization methods, the quadratic optimization problem over integers in~\eqref{l2-norm-min} must be reformulated as a Quadratic Unconstrained Binary Optimization (QUBO) problem defined over binary variables.
This reformulation enables the problem to be expressed in a form compatible with current quantum optimization hardware.

QUBO models play a central role in combinatorial optimization and have been applied across a wide range of domains~\cite{kochenberger2014unconstrained}.
In the context of quantum optimization, the QUBO formulation is particularly significant due to its computational equivalence to the Ising model from statistical physics~\cite{lucas2014ising, glover2019quantum}.
This correspondence allows discrete optimization problems to be mapped directly onto physical quantum systems, where the ground state of the associated Hamiltonian encodes the solution to the original problem.

In general, consider the set of all possible binary vectors $\boldsymbol{x} \in \{0,1\}^n$, with $n$ being the number of bits.
The function $f_Q: \{0, 1\}^n \rightarrow \mathbb{R}$ maps each binary vector $\boldsymbol{x}$ to a real value by:
    \begin{equation}
        \label{qubo}
        f_Q(\boldsymbol{x}) = \boldsymbol{x}^T Q \boldsymbol{x} = \sum_{i=1}^n \sum_{j=1}^n x_i Q_{ij} x_j \, .
    \end{equation}
The squared matrix $Q$ contains real-valued coefficients $Q_{ij}$ defining the interaction between variable $x_i$ and $x_j$.
The optimal solution of the corresponding QUBO problem is the binary vector~$\boldsymbol{x}^*$ minimizing the function $f_Q$:
    \begin{equation}
        \boldsymbol{x}^* = \arg \min_{\boldsymbol{x}} f_Q(\boldsymbol{x}) \, .
    \end{equation}
It is straightforward to show that the unfolding formulation in \eqref{l2-norm-min} is mathematically equivalent to the following optimization problem over the integer-valued vector $\boldsymbol{z}$:
    \begin{align}
        \label{int-optimization}
        \min_{\boldsymbol{z}} \left( \boldsymbol{a}^T \boldsymbol{z} + \boldsymbol{z}^T B \, \boldsymbol{z} \right)
        \quad\quad \mathrm{with} \quad\quad
            \begin{aligned}
                \boldsymbol{a} &= -2 {R}^T \boldsymbol{n} \, , \\
                B &= R^T R + \lambda D^T D \, .
            \end{aligned}
    \end{align}
However, to rewrite this problem as a QUBO in~\eqref{qubo}, each integer variable $z_i$ (i.e., the number of entries in bin $i$) must be encoded into a binary vector~$\boldsymbol{x}_i$ (i.e., a bitstring of length~$l_i$).
Following the approach in~\cite{krakoff_controlled_2022}, we use a generalized version of the proposed encoding strategy to have a tunable resolution in each bin.
In particular, each variable is encoded as $z_i = \boldsymbol{p}_i \cdot \boldsymbol{x}_i$, where $\boldsymbol{p}_i = (2^0, 2^1, \dots, 2^{l_i-1})$ is the precision vector of length $l_i$ for bin $i$.

The integer $l_i$ corresponds to the number of bits used in the representation of $z_i$, and thus sets the resolution.
It is determined as $l_i = \left\lceil \log_2 (\eta n_i / \epsilon_i) \right\rceil$, where $\epsilon_i$ is the bin by bin detection efficiency, $\eta$ is a hyperparameter controlling the desired resolution, and $\lceil q \rceil$ denotes the ceiling function that returns the smallest integer greater or equal to~$q$.
Using this encoding strategy, the linear and quadratic terms of the optimization problem in~\eqref{int-optimization} can be expressed as:
    \begin{align}
    \begin{aligned}
        \boldsymbol{a}^T \boldsymbol{z} &=
            \sum_{i=1}^M a_i \, \boldsymbol{p}_i \cdot \boldsymbol{x}_i =
            \boldsymbol{a}_{\mathrm{bin}}^T \, \boldsymbol{x} \, ,
        \\
        \boldsymbol{z}^T B \boldsymbol{z} &=
            \sum_{i=1}^M \sum_{i=j}^M \boldsymbol{x}_i^T \, \boldsymbol{p}_i^T \, B_{ij} \, \boldsymbol{p}_j \, \boldsymbol{x}_j =
            \boldsymbol{x}^T B_{\mathrm{bin}} \, \boldsymbol{x} \, .
            \label{eqn:discrete_rewriting}
    \end{aligned}
    \end{align}
The bitstring $\boldsymbol{x} \in \{0,1\}^n$ corresponds to the whole vector of binary variables, and it is obtained by concatenating all the bitstrings $\boldsymbol{x}_i$ for each bin.
The real-valued vector $\boldsymbol{a}_{\mathrm{bin}}$ and matrix $B_{\mathrm{bin}}$ represent respectively the linear and the quadratic interactions of the variables defined by the encoding procedure.
Notice that the scaling of the required total number of bits $n = \sum_{i=1}^M l_i$ is linear with the number of bins of the histogram and logarithmic with the number of entries in each bin.
Based on~\eqref{eqn:discrete_rewriting}, we can rewrite~\eqref{int-optimization} as the following minimization problem:
\begin{equation}
    \label{model}
    \min_{\boldsymbol{x}} \left( \boldsymbol{a}_{\mathrm{bin}}^T \, \boldsymbol{x} + \boldsymbol{x}^T B_{\mathrm{bin}} \, \boldsymbol{x} \right) \, .
\end{equation}
Since $x_i \in \{0,1\}$ such that $x_i^2 = x_i \, \forall i$, the expression of the objective function to be minimized in~\eqref{model} is actually equivalent to the usual compact form of the QUBO problem function in \eqref{qubo}.

\section{Results}\label{sec:results}
We evaluate the proposed optimization-based unfolding on four representative distributions commonly encountered in High-Energy Physics analyses: Normal, Exponential, Gamma, and Breit-Wigner.
These Probability Density Functions (PDFs) were selected to probe different structural features such as symmetry, strong skewness, heavy tails, and narrow resonant peaks.

\begin{figure}[!h]
\centering
    \begin{subfigure}[t]{0.49\textwidth}
    \centering
        \includegraphics[width=\textwidth]{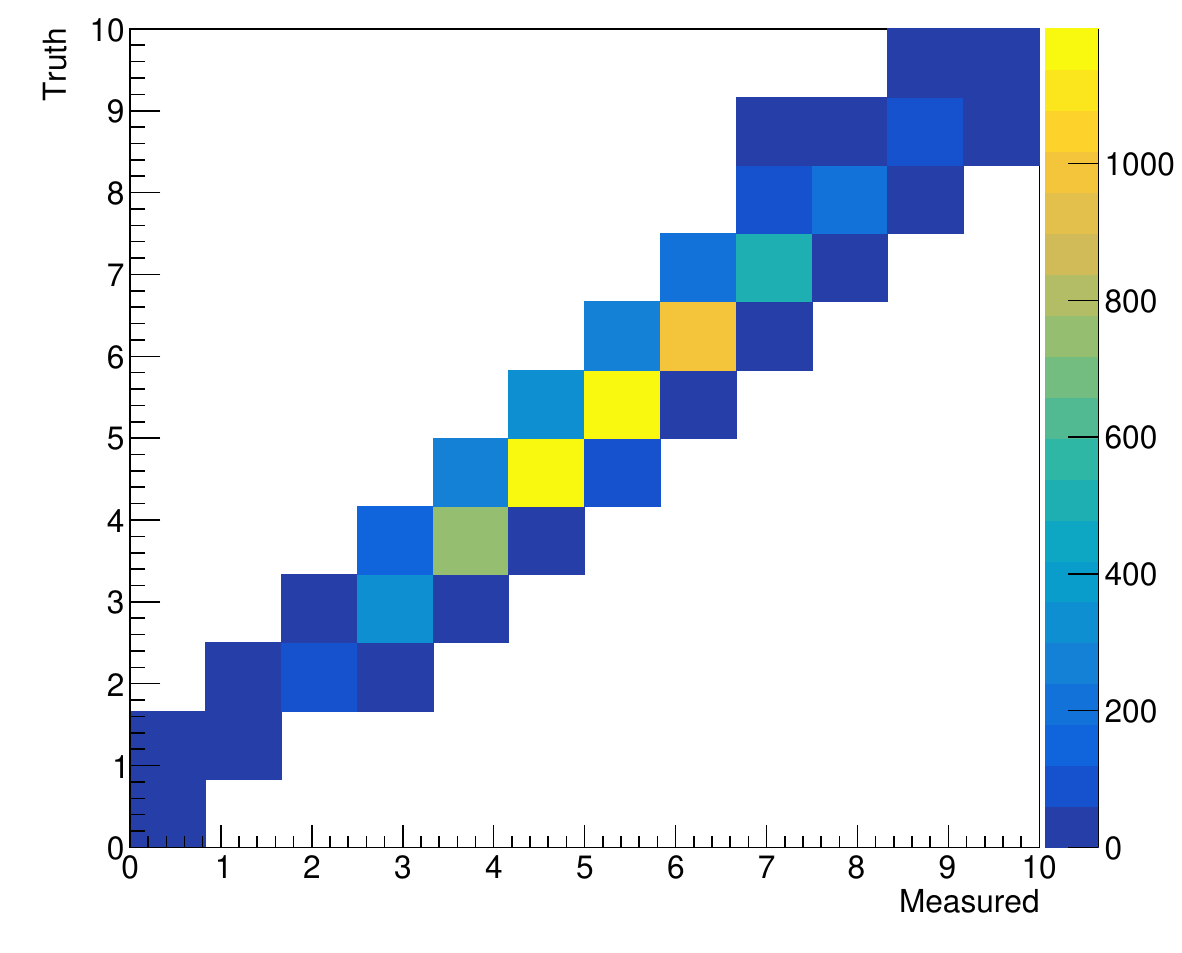}
        \caption{Normal distribution.}
    \end{subfigure}\hfill
    \begin{subfigure}[t]{0.49\textwidth}
    \centering
        \includegraphics[width=\textwidth]{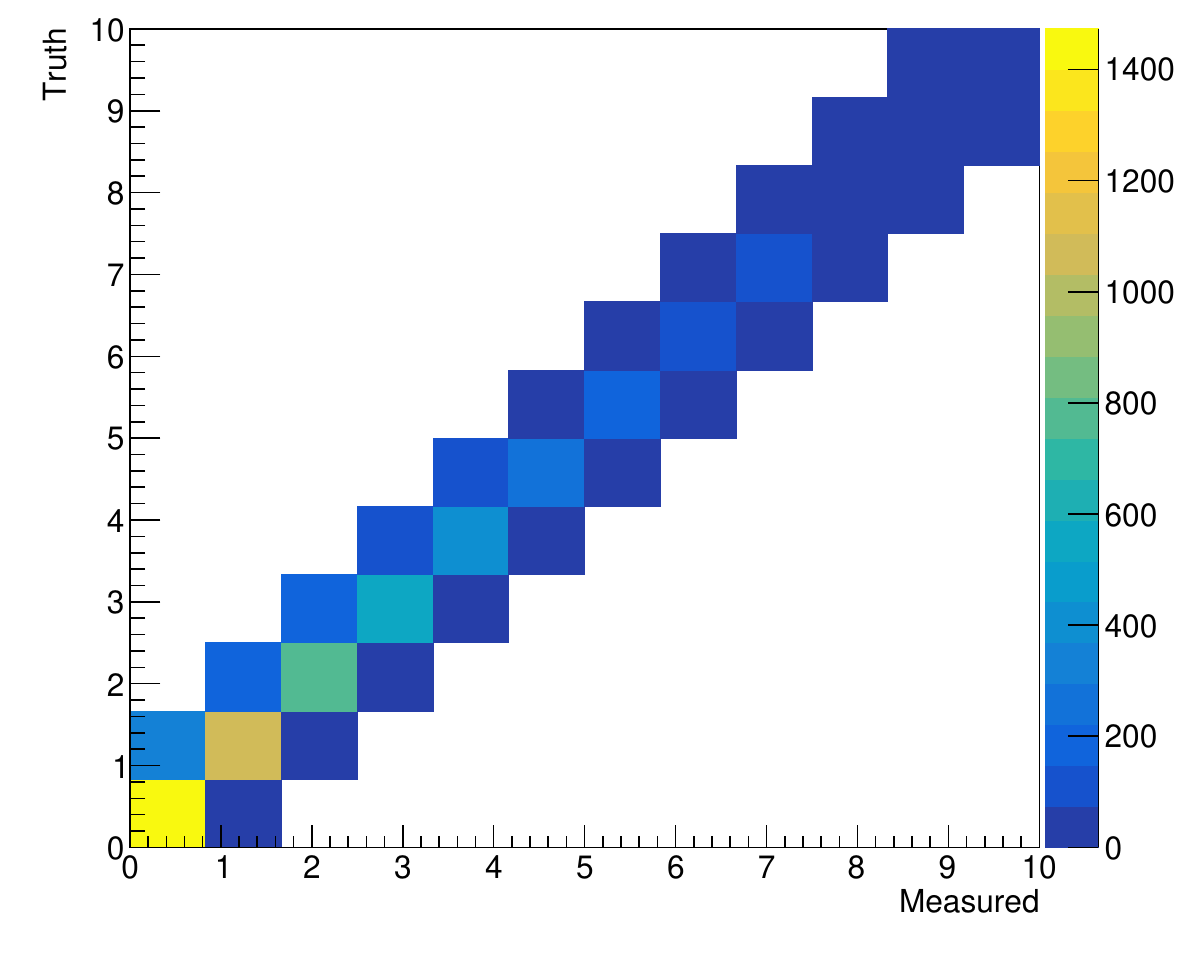}
        \caption{Exponential distribution.}
    \end{subfigure}\vspace{5mm}
    \begin{subfigure}[t]{0.49\textwidth}
    \centering
        \includegraphics[width=\textwidth]{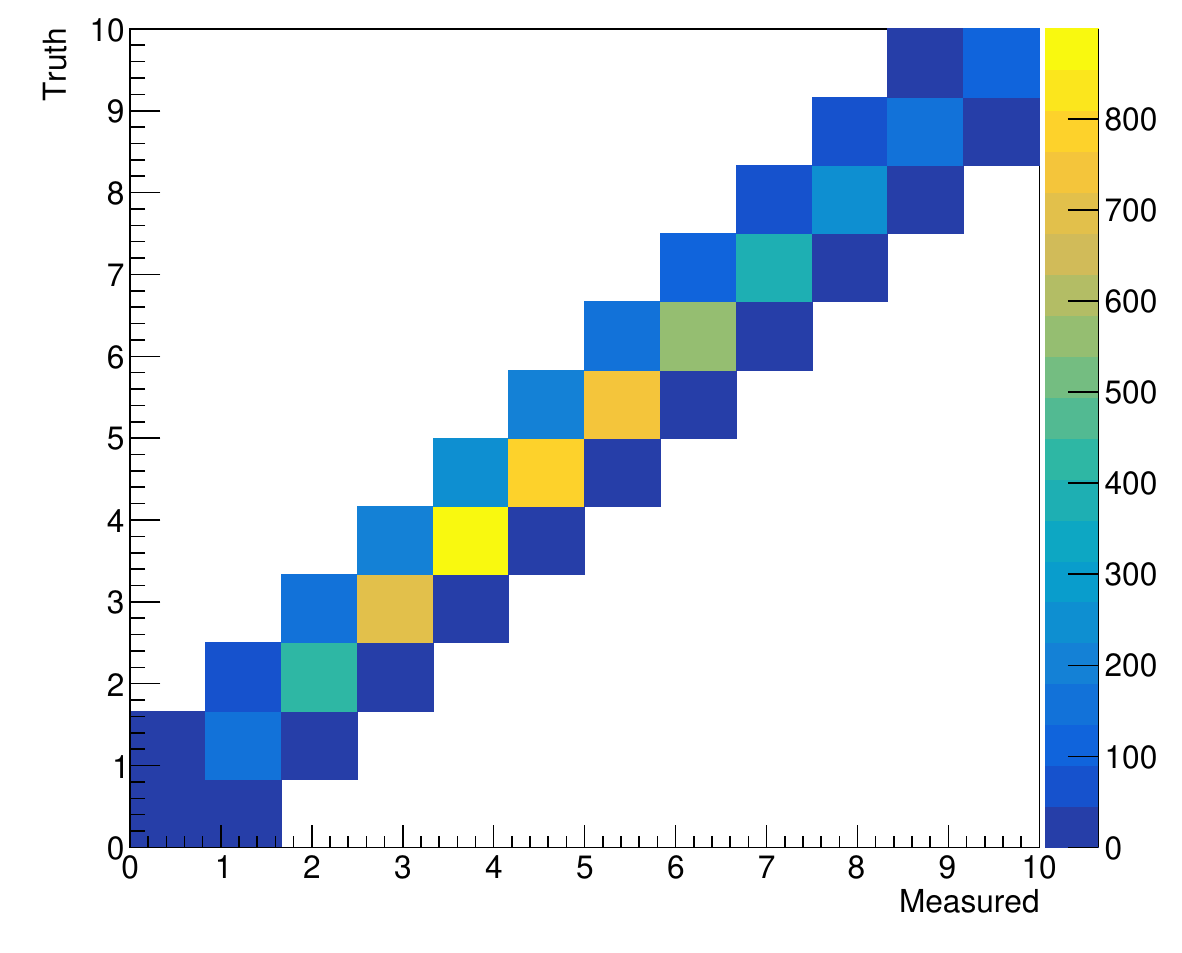}
        \caption{Gamma distribution.}
    \end{subfigure}\hfill
    \begin{subfigure}[t]{0.49\textwidth}
    \centering
        \includegraphics[width=\textwidth]{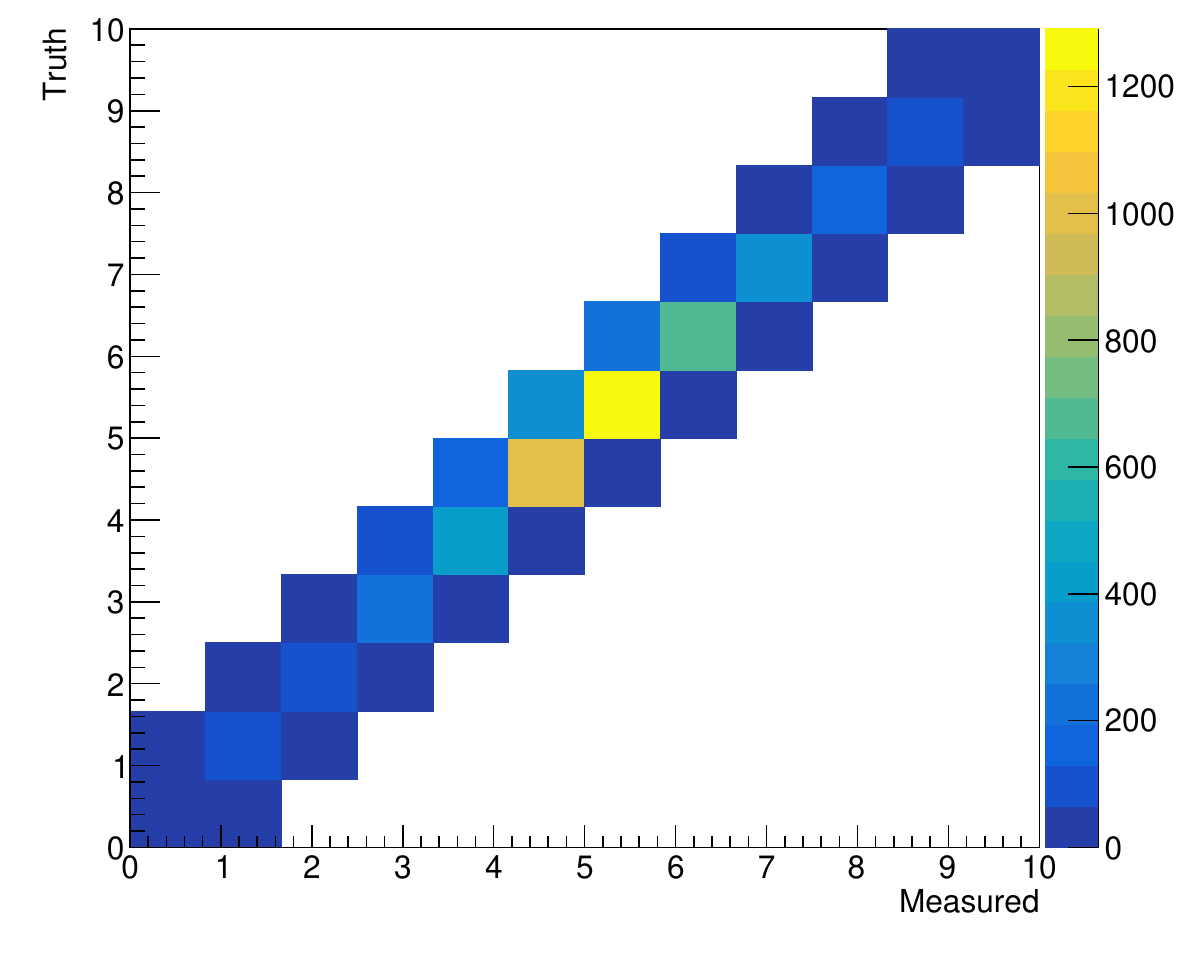}
        \caption{Breit-Wigner distribution.}
    \end{subfigure}
    \caption{Response matrices constructed for the different underlying distributions.
    The reconstructed bins are shown on the x-axis, while the truth-level bins are shown on the y-axis.
    The color scale indicates the number of truth-reco event pairs in each bin.}
    \label{fig:resp_matrices}
\end{figure}

For each distribution, two independent datasets of $10\,000$ entries distributed over $12$ equally spaced bins are artificially generated by sampling from the same underlying PDF.
The first dataset is used to construct the truth-reco pairs entering the response matrices shown in Fig.~\ref{fig:resp_matrices}, while the second dataset is used to generate the truth and measured histograms employed in the unfolding procedure.
Consequently, the truth distribution used to build the response matrix and truth distribution associated with the measured data correspond to different statistical realizations of the same PDF.
This setup therefore assumes a perfectly modeled simulation, where the response matrix accurately reproduces the detector response of the data under study.
The goal of the present analysis is thus not to investigate the resilience of the optimization-based unfolding framework against possible mismodeling or data-simulation discrepancies, but rather to assess the intrinsic performance of the proposed optimization approach under controlled conditions.
Detector effects are modeled through a response matrix incorporating Gaussian smearing, a fixed bias shift, and limited detection efficiency.
In particular, detector smearing and reconstruction bias are simulated by adding to each generated event a Gaussian random variable sampled from $\mathcal{N}(\mu, \sigma)$ with $\mu=-0.08$ and $\sigma=0.17$, while the detection efficiency is fixed to $\epsilon = 0.8$ uniformly across all bins.

We compare five unfolding approaches:
\begin{itemize}
\item Response Matrix Inversion (MI),
\item Iterative Bayesian Unfolding (IBU),
\item Singular Value Decomposition (SVD) unfolding,
\item Quadratic integer optimization solved with Gurobi (GRB),
\item QUBO solved with the D-Wave hybrid quantum-classical solver (HYB).
\end{itemize}

For all regularized methods, hyperparameters are not individually optimized for each distribution. Instead, we adopt fixed values chosen to provide stable solutions while limiting overfitting effects.
In particular, the number of iterations used for IBU is fixed to $4$, while the regularization strength employed in the optimization-based approaches is set to $\lambda = 0.1$.
Although no dedicated hyperparameter tuning procedure is performed, we empirically observe that the unfolding results remain stable and qualitatively robust under moderate variations of the regularization strength.

To quantify the agreement between the unfolded estimate and the reference truth histogram, we use a Pearson-like bin-wise discrepancy measure, defined by:
\begin{equation}
    \label{chi2}
    \chi^2 = \sum_{i=1}^{M} 
    \frac{\left(z_i^{\text{true}} - \hat{z}_i \right)^2}{z_i^{\text{true}}}\,.
\end{equation}
We emphasize that this quantity is used here only as a reconstruction quality metric for comparing unfolding methods under controlled synthetic conditions.
It does not include the covariance matrix of the unfolded estimate, nor the statistical uncertainties associated with the finite sample used to construct the response matrix.

\begin{figure}[!h]
\centering
    \begin{subfigure}[t]{0.49\textwidth}
    \centering
        \includegraphics[width=\textwidth]{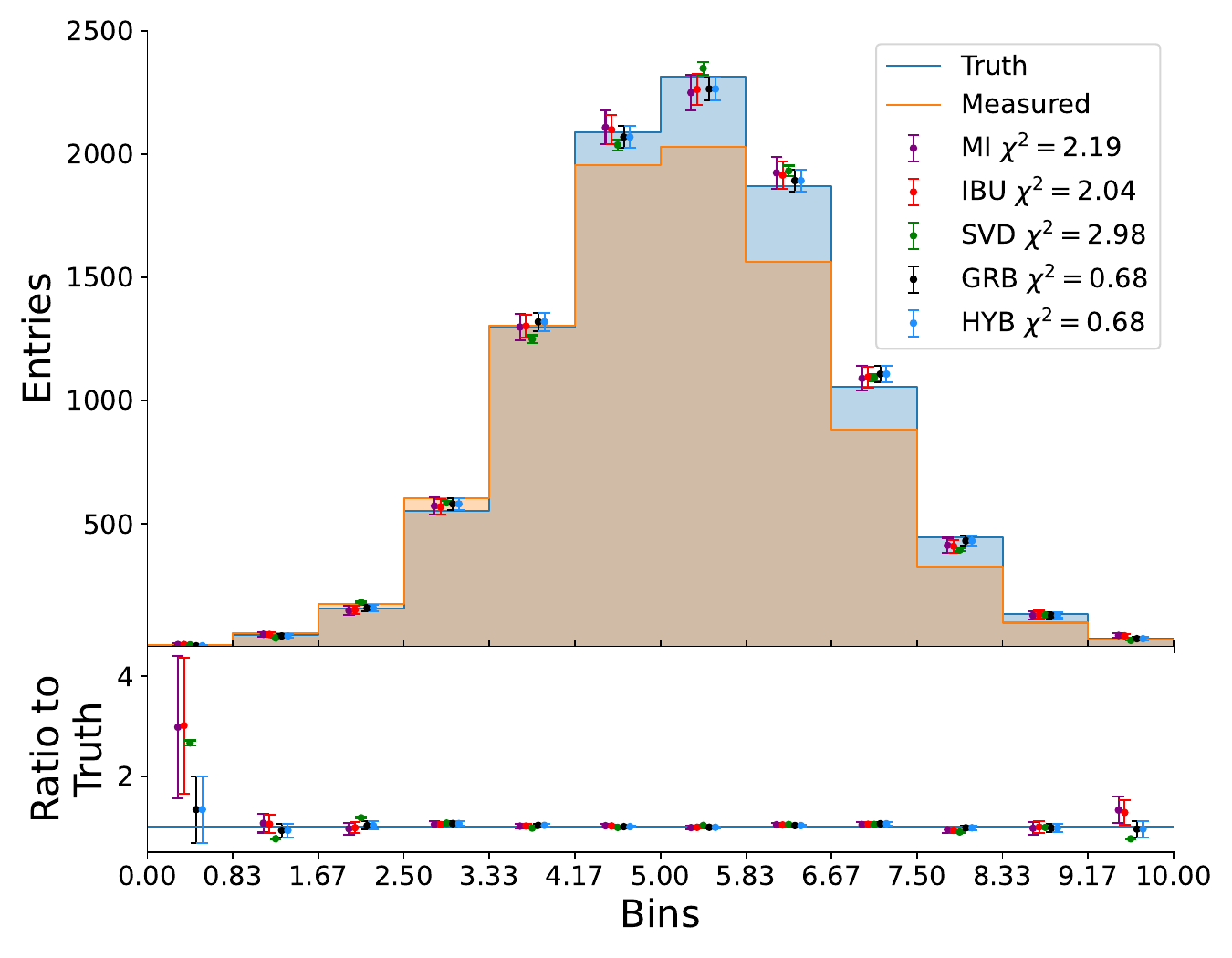}
        \caption{Normal distribution.}
    \end{subfigure}\hfill
    \begin{subfigure}[t]{0.49\textwidth}
    \centering
        \includegraphics[width=\textwidth]{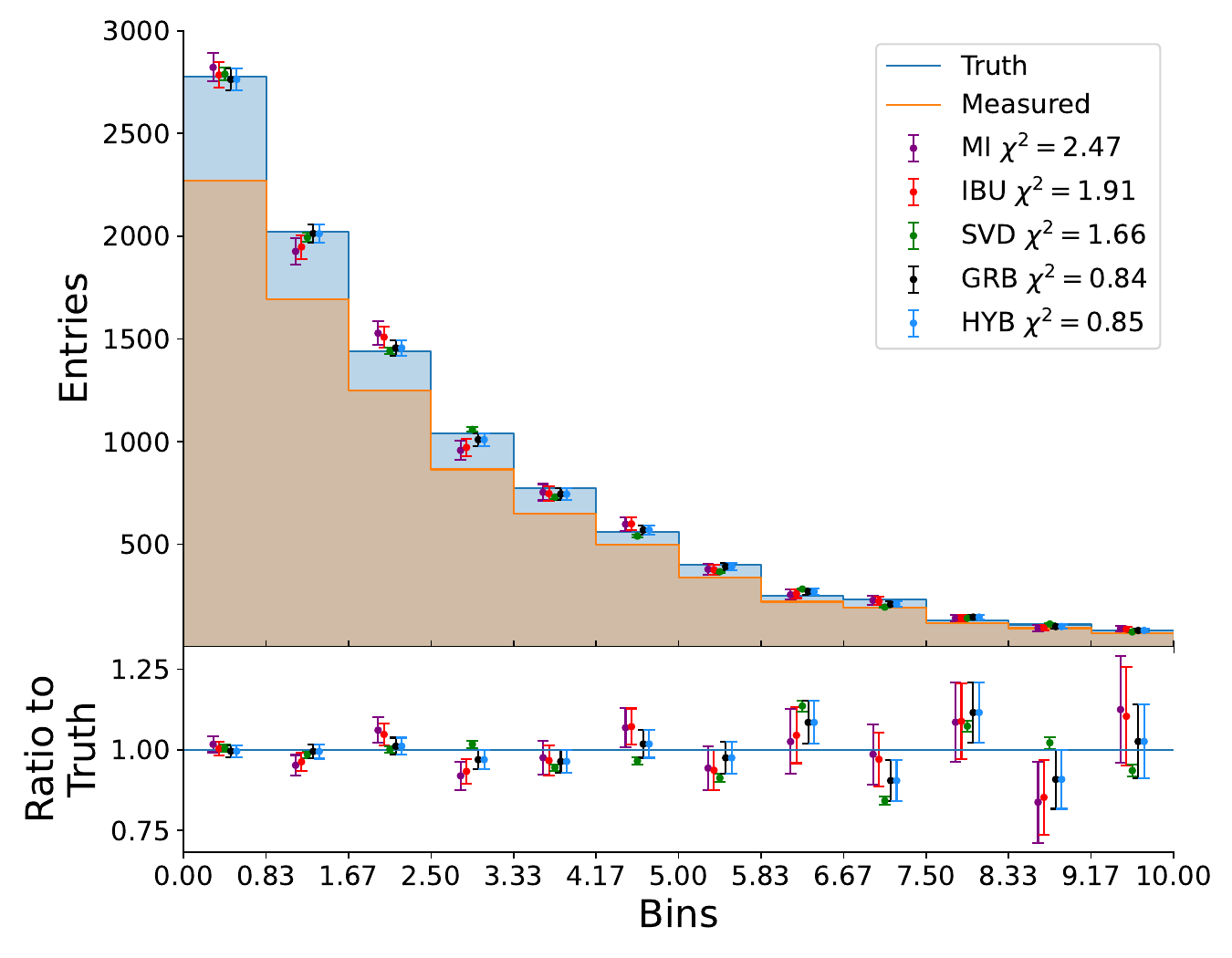}
        \caption{Exponential distribution.}
    \end{subfigure}\vspace{5mm}
    \begin{subfigure}[t]{0.49\textwidth}
    \centering
        \includegraphics[width=\textwidth]{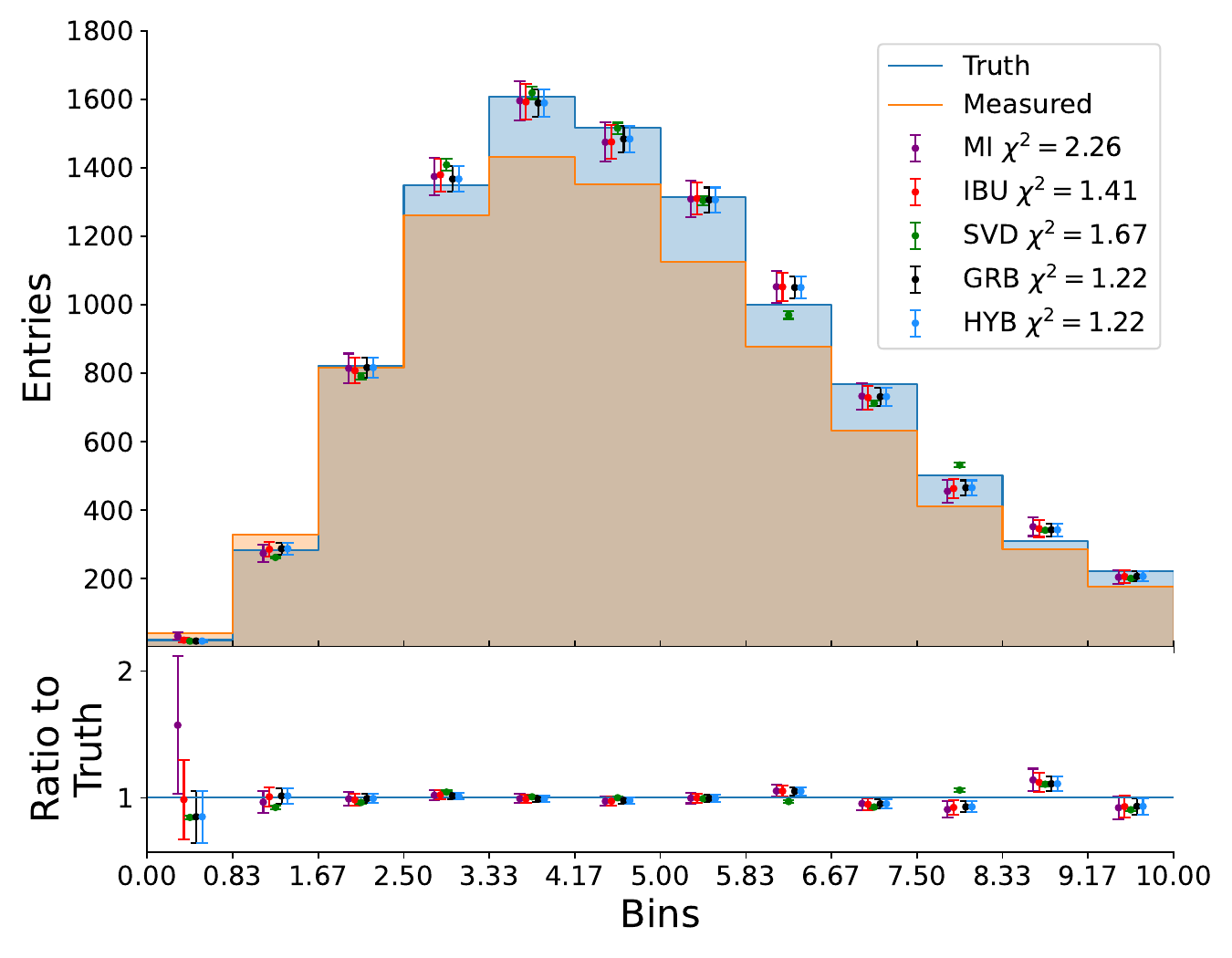}
        \caption{Gamma distribution.}
    \end{subfigure}\hfill
    \begin{subfigure}[t]{0.49\textwidth}
    \centering
        \includegraphics[width=\textwidth]{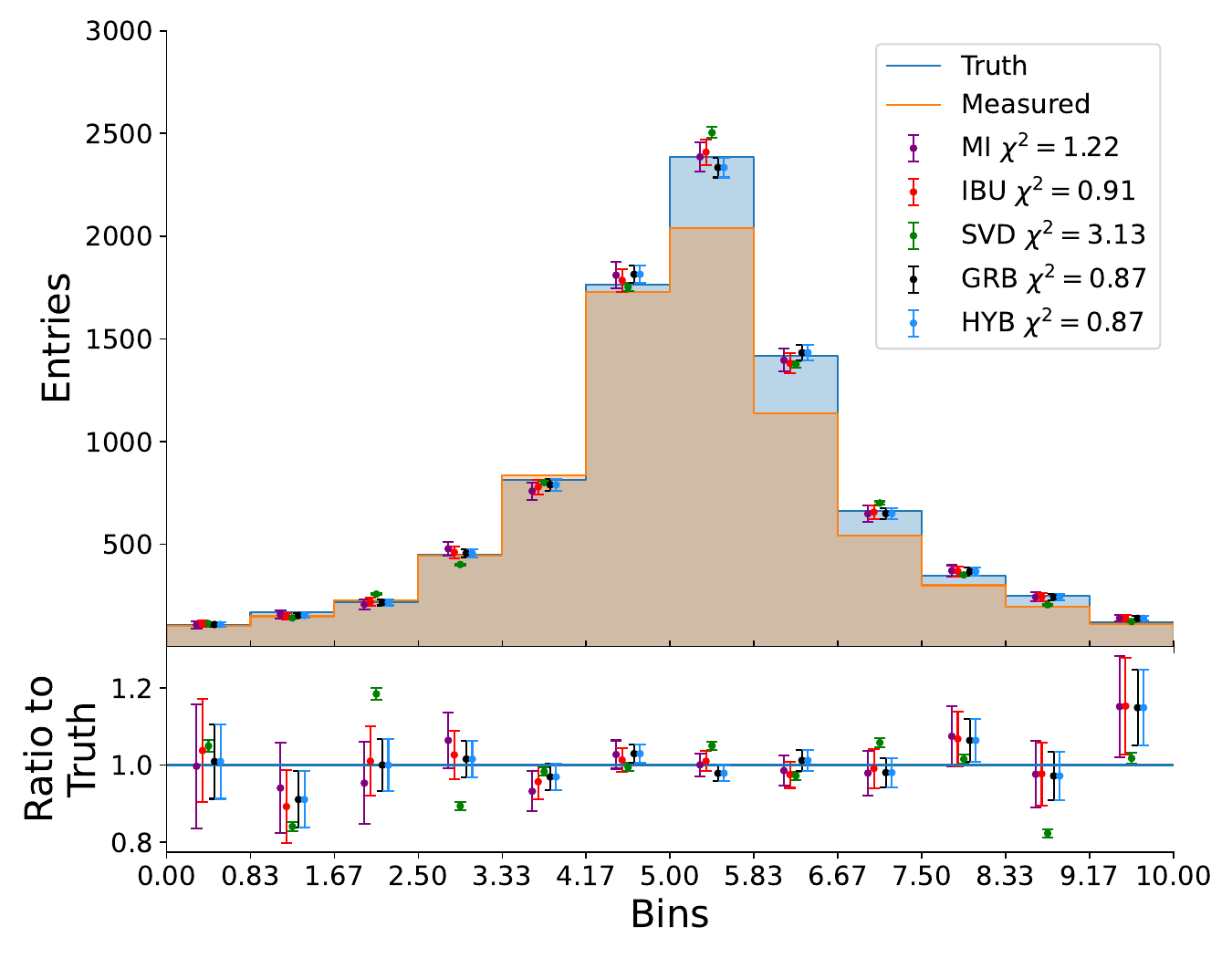}
        \caption{Breit-Wigner distribution.}
    \end{subfigure}
    \caption{Comparison of results obtained with different unfolding methods.
    True histograms are shown in blue, measured histograms in orange, and unfolded results as colored markers with statistical uncertainties.
    The methods compared are Matrix Inversion (MI), Iterative Bayesian Unfolding (IBU), Singular Value Decomposition (SVD), classical optimization using Gurobi (GRB), and hybrid quantum-classical solver (HYB).
    The values reported in the legends correspond to the Pearson-like discrepancy metric used for relative comparison among methods.}
    \label{fig:res_histograms}
\end{figure}

In Fig.~\ref{fig:res_histograms}, we present the unfolding results obtained with the different methods for each synthetic benchmark distribution.
The lower panels show the bin-wise ratio $\hat{z}_i / z_i^{\text{true}}$, providing a direct visualization of local deviations and bias patterns.
Here, $z_i^{\text{true}}$ denotes the finite truth histogram associated with the synthetic benchmark sample used for the unfolding test, rather than the exact analytic probability density from which the events are generated.

The Matrix Inversion (MI) method exhibits oscillatory behavior in low-statistics bins, highlighting its sensitivity to statistical fluctuations and the ill-conditioned nature of the response matrix.
Regularized approaches such as SVD and IBU improve stability, but residual deviations remain visible, particularly in regions with steep gradients and localized structures.
The optimization-based approaches (GRB and HYB) achieve the lowest values of the Pearson-like discrepancy metric and maintain bin-wise ratios close to unity over most of the spectrum.
Across symmetric, skewed, heavy-tailed, and resonant shapes, the optimization-based solutions demonstrate stable behavior in both central and low-occupancy regions.
In particular, they balance smoothness and fidelity effectively, mitigating oscillations while preventing the over-smoothing that can arise in aggressively regularized solutions.
The hybrid quantum-classical solver reproduces the classical optimization results with high consistency, indicating that the QUBO reformulation preserves the structure of the original quadratic optimization objective.
Taken together, these results suggest that the proposed optimization framework provides competitive performance across different distributional features.

The primary implication is therefore methodological: unfolding can be formulated in a way that is compatible with both classical and quantum-enabled optimization paradigms.
These findings support the viability of treating unfolding as a discrete optimization problem and provide a foundation for further investigations into scalability, hyperparameter selection, and applications to more realistic experimental datasets.

\section{Conclusion}\label{sec:conclusion}
In this work, we introduced an optimization-based framework for statistical unfolding in High-Energy Physics.
Starting from a regularized likelihood formulation, we derived an equivalent quadratic integer optimization problem that naturally incorporates detector response effects and smoothness constraints.
This formulation enables the use of modern optimization techniques, including mixed-integer quadratic programming and QUBO-based solvers compatible with quantum annealing hardware.
The proposed approach was implemented in the open-source \texttt{QUnfold} package, providing a flexible interface to classical and hybrid quantum-classical solvers.

We benchmarked the method against widely used unfolding techniques, namely Matrix Inversion, Iterative Bayesian Unfolding, and Singular Value Decomposition, using four representative synthetic distributions with controlled detector distortions.
Across all considered scenarios, the optimization-based approach achieved reconstruction quality competitive with established methods in terms of the Pearson-like discrepancy metric and bin-wise stability.
In particular, the method demonstrated robust behavior in regions with steep gradients, localized resonant structures, and low-statistics bins.
A key outcome of this study is that the QUBO reformulation preserves the structure and performance of the original quadratic objective.
The hybrid quantum-classical solver produced solutions consistent with those obtained using a state-of-the-art classical mixed-integer solver.
These results establish the methodological feasibility of mapping unfolding problems onto quantum-compatible optimization frameworks, providing a concrete pathway for future investigations as quantum hardware continues to mature.

From a broader perspective, treating unfolding as a discrete optimization problem offers several conceptual and practical benefits.
Regularization is naturally embedded in the objective function, constraints can be incorporated explicitly, and the formulation is compatible with a wide class of optimization strategies.
Moreover, the integer-based representation allows direct control over bin resolution and precision, which may be advantageous in analyses with discrete event counts or limited statistics.

Several directions for future work remain.
First, systematic studies based on repeated pseudo-experiments are required to quantify bias-variance trade-offs and statistical robustness in realistic scenarios, including the covariance of the unfolded distributions and the statistical uncertainties on the response matrices.
Second, scalability tests with larger bin numbers and higher-dimensional observables will be essential to assess performance in real HEP analysis scenarios.
Finally, the integration of more sophisticated regularization schemes or data-driven hyperparameter selection strategies could further improve reconstruction quality.

\section*{Acknowledgments}
SG and GB thank R. Di Sipio for fruitful discussions and CINECA for providing access to the D-Wave quantum resources.
CR is sponsored by the Wolfgang Gentner Programme of the German Federal Ministry of Education and Research (grant no. 13E18CHA).
MG and CR are supported through the CERN Quantum Technology Initiative.

\bibliographystyle{unsrt}
\bibliography{references}

\end{document}